\documentclass[aps,prl,twocolumn,showpacs,groupedaddress]{revtex4}

\usepackage{amsmath}
\usepackage{graphicx}
\usepackage{dcolumn}
\usepackage{bm}

\begin{document}


\title{Ground-state energy and stability limit of small $^{\bm 3}$He drops}

\author{E. Sola}
\author{J. Casulleras}
\author{J. Boronat}
\affiliation{Departament de F\'\i sica i Enginyeria Nuclear, Campus Nord
B4-B5, Universitat Polit\`ecnica de Catalunya, E-08034 Barcelona, Spain}

\date{\today}
\begin{abstract}
Small and stable drops of $^3$He atoms can only exist
above a minimum number of particles, due to the combination
of the $^3$He atom Fermi statistics and its light mass. An 
accurate estimation of this minimum number using microscopic
theory has been difficult due to the inhomogeneous and fermionic
nature of these systems. We present a diffusion Monte Carlo 
calculation of $^3$He drops with sizes near the minimum   
in order to determine the stability threshold. The results show 
that the minimum self-bound drop is formed by $N=30$
atoms with preferred orbitals for open shells corresponding
to maximum value of the spin.
\end{abstract}

\pacs{67.55.-s,36.40.-c,02.70.Ss}

\maketitle

Liquid $^3$He drops offer a unique combination in nature of Fermi
statistics, neutral charge and self-bound character~\cite{jcp,barranco_rev}. The
interest in their physical knowledge explains the continued effort from
both the experimental~\cite{toennies} and  theoretical
sides~\cite{stringari,pandha,barranco,guardio_vmc,guardio_dmc} towards a
better description of their particular properties. Experimentally, $^3$He
drops are currently generated in the laboratory by means of a free jet
expansion from a stagnation source chamber through a thin walled
nozzle~\cite{jcp}.  The estimated temperature of the Fermi drops is 0.15 K
and therefore they are in their normal state. The non-superfluid character
of $^3$He drops has been detected in a series of
experiments~\cite{superfluid} where single molecules were embedded in the
drops.  If the molecule is surrounded by $^4$He, the rotational spectrum
presents a sharp structure which has been attributed to the superfluid
nature of $^4$He, whereas for a $^3$He drop a broad peak is observed.

The smaller mass of the $^3$He atom, and more fundamentally its Fermi
statistics, introduces sizeable differences in the stability of the drop
with respect to $^4$He. Two $^4$He atoms form a bound state but a  minimum
number of $^3$He atoms is necessary to form a self-bound
system~\cite{pandha}. This difference is observed in the experimental setup
since very small $^4$He drops are detected in the jet, starting from the
dimer, whereas it  has been proven to be difficult to observe $^3$He drops
with less than 1000 atoms~\cite{toennies}. Reducing this number and 
generating drops with a number of atoms closer to the  threshold limit is one
of the most important challenges for the next future. Small $^3$He drops
are expected to present ``magic numbers"~\cite{stringari}  whose
experimental  determination would be a clear signature of their Fermi
statistics and a new benchmark for quantum many-body theories for
inhomogeneous systems.  

As a result of the complexity arising from the combination of Fermi 
statistics and inhomogeneity,
the number of microscopic works on $^3$He drops is significantly 
smaller than those devoted to $^4$He drops.
The first systematic study
was carried out by Pandharipande  \textit{et al.}~\cite{pandha} in the
eighties  using the variational Monte Carlo (VMC) method. This calculation
used a trial wave function incorporating backflow correlations to correct
the nodal surface of the noninteracting system and predicted that a drop
with 40 atoms is self bound. More recently, Guardiola and
Navarro~\cite{guardio_vmc} carried out a detailed VMC calculation of small
$^3$He drops including in the trial wave function
configuration-interaction-like correlations. These new correlations
improved the energy in a significant way and the smallest bound drop was
estimated to be the one formed by 35 atoms. Recently, the same
authors~\cite{guardio_dmc} have obtained a lower value (32) for this
upper-bound  threshold  in a diffusion Monte Carlo calculation (DMC)
restricted to the atom's number range 31-34. Therefore, the improvement of
both the trial wave function and the theoretical approach has progressively
reduced the minimum number of atoms required for a self-bound drop. It is
worth mentioning that using a non-local density-functional
approach~\cite{barranco}  this critical number was  estimated to be
slightly smaller (29) than these microscopic calculations.

In this work, we present a DMC calculation of small $^3$He drops around
the threshold limit for self-binding. In the simulation we use the
fixed-node (FN) approximation~\cite{fixedn}, which provides an upper bound
to the exact eigenvalue, and the release-node (RN) method~\cite{releasedn}
to estimate the quality of the FN upper-bound. The approach is the same we
followed in the past in the DMC calculation of the equation of state of
bulk~\cite{casu} and two-dimensional~\cite{grau} $^3$He. The results show
that the minimum number for a self-bound drop is 30 and that in open-shell
configurations the optimal energy corresponds to maximum spin. The latter
conclusion was also pointed out in previous density
functional~\cite{barranco} and VMC work~\cite{guardio_vmc}.

The sign problem in a  DMC simulation is usually dealt   within the FN
approximation~\cite{fixedn}. Along the calculation, the wave function
$f({\bf R},\tau)=\psi_{\text T}({\bf R}) \Psi({\bf R},\tau)$
(${\bf R}={\bf r}_1,\ldots,{\bf r}_N$) evolves according to the imaginary-time
($\tau$) Schr\"odinger equation, with $\psi_{\text T}$ acting as
importance sampling function and nodal constraint. For long enough time
$\tau \rightarrow \infty$, $\Psi({\bf R})$ is the lowest energy state
compatible with the nodal surface imposed by $ \psi_{\text T}({\bf R})$.  
The trial wave function we have used has a Jastrow-Slater form,
\begin{equation}
\psi_{\text T}({\bf R}) = \prod_{i<j}^{N} f(r_{ij}) \, \Phi_\uparrow
\Phi_\downarrow  \ ,
\label{trial1}
\end{equation}
with a two-body correlation factor,
\begin{equation}
f(r) = \exp \left[ -\frac{1}{2} \left( \frac{(\alpha r)^\beta}{N} + \left(
\frac{b}{r} \right)^\nu \right) \right]  \ .
\label{jastrow}
\end{equation}
In Eq. (\ref{trial1}), $\Phi_\uparrow$ ($\Phi_\downarrow$) is  a Slater
determinant for the spin-up (spin-down) particles filled up with
single-particle orbitals corresponding to the polynomial part of the
harmonic oscillator basis. From a practical point of view, the use of this
basis is clearly advantageous since the resulting Slater determinant is of
the Vandermonde type and therefore translationally invariant. With this
model, and considering the Jastrow part as a function of relative distances
only, spurious energy contributions due to the movement of the  center of
mass are not present. The orbitals are chosen in its Cartesian coordinate
representation, and for incomplete shells we have followed the prescription
used by Guardiola and Navarro~\cite{guardio_vmc} which warrants invariance
under $90^0$ rotations with respect to the Cartesian axis. 

The \textit{quality} of the upper bound in the FN approach depends on
the accuracy of the nodal surface defined by $\psi_{\text T}({\bf R})$. 
The nodal surface of  $\psi_{\text T}({\bf R})$ (\ref{trial1}) corresponds
to the one of a non-interacting system. Therefore, this model is expected
to be too simple for describing a correlated liquid as $^3$He. Using the
imaginary-time Schr\"odinger equation, one can prove that the first
correction (corresponding to a short imaginary-time interval) to the
non-interacting nodal surface corresponds to a displacement of the
coordinates due to correlations with all the other
particles~\cite{boro_review}. These corrections are known as backflow
correlations and are constructed by replacing the coordinates ${\bf r}_i$
of particles in the Slater determinants by 
\begin{equation}
\tilde{\bf r}_i = {\bf r}_i + \lambda_{\text B} 
\sum_{j \neq i}^{N} \eta(r_{ij}) {\bf r}_{ij}    \ .
\label{backf}
\end{equation}  
Similarly to previous studies of the homogeneous liquid~\cite{casu,grau}, 
we have used for the function $\eta(r)$ a
Gaussian, $\eta(r)=\exp[{-((r-r_{\text B})/\omega_{\text
B})^2}]$; $\lambda_{\text B}$, $r_{\text B}$, and $\omega_{\text
B}$ are variational parameters. 

The FN method provides a rigorous upper-bound on the ground-state energy
but does not provide information on the quality of the upper bound, i.e.,
the difference between the energy obtained and the exact eigenvalue. In
order to estimate the bias due to the particular model nodal surface
we have used the released node (RN) technique~\cite{releasedn}. In
the RN method, the walkers are allowed to cross the nodal surface
determined by  $\psi_{\text T}({\bf R})$ for a finite lifetime $t_{\text
r}$ and a sign $+$ or $-$ is assigned to each one. To this end, the
importance sampling wave function is chosen positive
\begin{equation}
\psi({\bf R}) = \left( \psi_{\text T}({\bf R})^2 + a^2 \right)^{1/2} \ ,
\label{release}
\end{equation} 
with $a$ a constant, and the fermionic energy is obtained by projecting on
the antisymmetric component $\psi_{\text T}({\bf R})$. The method would
arrive to the exact ground-state energy for $t_{\text r} \rightarrow
\infty$, but this limit is not accesible in liquid $^3$He due to the rapid
emergence of bosonic noise. Nevertheless, the initial slope can be well
determined and its value can be used for comparing different nodal surfaces
and for an estimate of the magnitude of the bias introduced by the FN
approximation~\cite{casu}.

\begin{table}
\centering
\begin{ruledtabular}
\begin{tabular}{ccdd}
$ N  $ & $   S_z  $ & \multicolumn{1}{c}{$\ \ \ \ \ \ \ \ \ E/N$ (K)} & \multicolumn{1}{c}
{$\ \ \ \ \ \ \ \ \ K/N$ (K)} \\
\hline \\[-0.1cm] 
 29  &  9/2 & 0.0194(10)   & 3.395(65)  \\[0.1cm]   
 30  &   5  & -0.0006(11)  & 3.630(13)  \\
     &   4  &  0.0067(12)  & 3.630(15)  \\
     &   3  &  0.0184(12)  & 3.595(13)  \\[0.1cm]
 31  &  9/2 & -0.0078(12)  & 3.682(18)  \\
     &  3/2 &  0.0056(12)  & 3.666(13)  \\[0.1cm]
 32  &   4  & -0.0258(18)  & 3.808(15)  \\   
     &   3  & -0.0180(12)  & 3.729(11)  \\
     &   0  &  0.0003(10)  & 3.751(12)  \\[0.1cm]
 33  &  7/2 & -0.0377(13)  & 3.822(13)  \\
     &  5/2 & -0.0340(11)  & 3.836(14)  \\
     &  1/2 & -0.0190(13)  & 3.857(15)  \\[0.1cm]
 34  &   3  & -0.0535(15)  & 3.935(16)  \\
     &   0  & -0.0330(12)  & 3.942(16)  \\[0.1cm]
 35  &  5/2 & -0.0649(13)  & 4.02(2)    \\   
     &  3/2 & -0.0639(16)  & 3.999(19)  \\[0.1cm]
 36  &   2  & -0.0839(15)  & 4.122(13)  \\
     &   1  & -0.0792(17)  & 4.124(18)  \\[0.1cm]
 37  &  3/2 & -0.1016(15)  & 4.228(17)  \\[0.1cm]
 38  &   1  & -0.1211(19)  & 4.26(3)    \\    
     &   0  & -0.118(2)    & 4.241(18)  \\[0.1cm]
 39  &  1/2 & -0.1372(16)  & 4.349(16)  \\[0.1cm]
 40  &   0  & -0.1564(17)  & 4.442(17)  \\[0.1cm]
 43  &  3/2 & -0.1702(15)  & 4.583(16)  \\[0.1cm]
 55  & 15/2 & -0.2848(18)  & 5.463(16)  \\[0.1cm]
 70  &   0  & -0.412(2)    & 5.72(2)    \\
\end{tabular}
\end{ruledtabular}
\caption{Total ($E$) and kinetic ($K$) energy per particle of small $^3$He
drops as a function of the number of atoms. $S_z$ is the $z$ component of
the total spin of the drop. Figures in parenthesis are the statistical
errors.}
\end{table}

All the DMC simulations have been carried out using the HFD-B(HE) Aziz
potential~\cite{aziz}, which has proved high accuracy in the microscopic
description of the bulk phases of liquid $^4$He and
$^3$He~\cite{boro_review}. The parameters entering into $\psi_{\text
T}(\bf{R})$, Eqs. (\ref{jastrow}) and (\ref{backf}), are adjusted
variationally. The dependence on the number of atoms in the drop is only
significant for the parameter $\alpha$ (\ref{jastrow}): for $N=30$,
$\alpha=4.0\ \sigma^{-1}$ and for $N=40$, $\alpha=4.4\ \sigma^{-1}$,
increasing linearly with $N$ ($\sigma=2.556$\ \AA). The optimal values of
the rest of parameters are: $\beta=1$, $\nu=5$,  $b=1.14\ \sigma$,
$\lambda_{\text B}=0.34$, $r_{\text B}=0.75\ \sigma$, and $\omega_{\text
B}=0.54 \ \sigma$.  

\begin{figure}
\centering
        \includegraphics[width=0.8\linewidth]{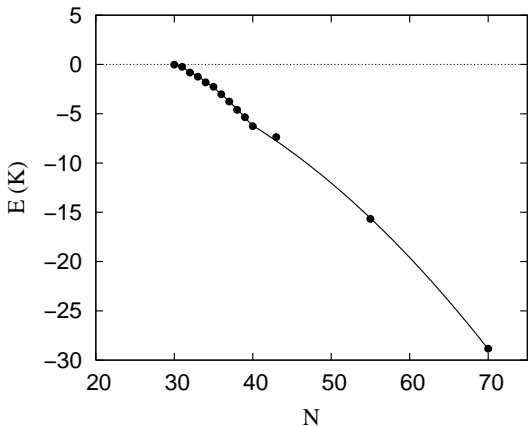}%
        \caption{Total energy of $^3$He drops as a function of the number
	of atoms $N$. The error bars are smaller than the symbol size. The
	line on top of the DMC data is a guide to the eye. }
\end{figure}

Table I contains results for the total ($E/N$) and kinetic energy ($K/N$)
per particle as a function of the number of atoms $N$ in the drop. With the
exception of the first row for $N=32$, 35, and 38 the calculations have
been made using orbitals with invariance under 90$^0$ rotations with
respect to the coordinate axis. According to our results, the threshold
limit for a self-bound drop is $N=30$ and all cases studied with equal $N$
but diferent spin $S_z$ show a preferred state corresponding to the
maximum value of the spin. The latter result can be also interpreted taking
into account the magic numbers which close a shell (of the spin-up or
spin-down atoms), which in the range studied correspond to values 10, 20,
and 35. The results contained in Table I show that the optimal energies
follow the rule of having at least one closed shell, with preference
for the smallest one: 10 for $N=29$ and 20 for $N=31-39$.

The lowest total energies of the $^3$He drops are shown in Fig. 1 as a
function of their number of atoms $N$. The line on top of the DMC data
correspond to polynomial fits and are only intended as to guide the eye.
As one can see in the figure, the behavior of the energy with $N$ is not
monotonous in the regime studied showing a kink for $N=40$, a doubly magic
number $N_\uparrow=N_\downarrow=20$. This kink is a remnant of the shell
model chosen to describe the antisymmetry of the system in absence of
correlations. Dynamical correlations induced by the interatomic
potential smooth significantly this effect but, for these small drops,
it is still clearly observable. On the other hand, Fig. 1 shows that in the
regime $N=30-40$ the DMC data display a regular behavior, which is well
reproduced by a second-degree polynomial. This feature yields us confidence
on the calculation itself and on the method followed for the selection of orbitals
in the Slater determinants.

\begin{figure}
\centering
        \includegraphics[width=0.8\linewidth]{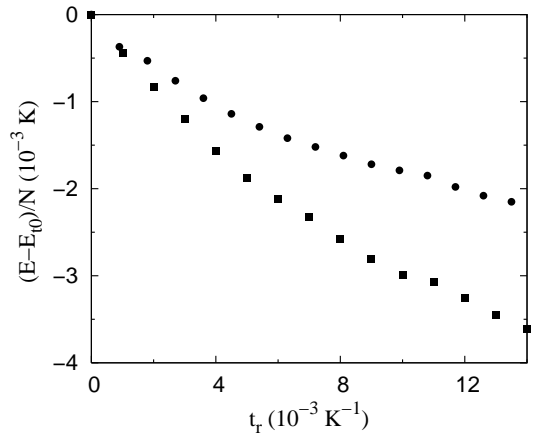}%
        \caption{Evolution of the energy with released time $t_{\text r}$ 
	for a $N=40$ $^3$He drop. Circles and squares correspond
	to Jastrow-Slater wave functions with and without
	backflow correlations, respectively. In both cases, we subtract to
	the RN energies the
	energy at initial time $E_{\text t0}=E(t_{\text r}=0)$ for an
	easier comparison. The statistical error bars are essentially
	constant in this $t_{\text r}$ range; they are not shown for
	major clarity of the released signal.}
\end{figure}

Although the search for an exact and stable quantum Monte Carlo 
algorithm for solving the  $N$-fermion Schr\"{o}dinger equation 
continues, the intrinsic difficulty of the problem raises the question 
about the maximum information one can obtain at present from 
the available Monte Carlo methods. The only really stable method, 
which can manage a significant number of fermions, is FN. With FN 
one is able of computing rigorous upper bounds to the ground-state 
energy, with the only constraint of the model nodal surface contained 
in  $\psi_{\text{T}}$. The introduction of backflow correlations in the 
model has proven to be of crucial importance in order to decrease the 
bias introduced by  $\psi_{\text{T}}$. Also for  $^3$He drops, this 
introduction allows for a much better description: for  $N=40$, backflow 
correlations make the energy per particle to decrease 0.03 K, 
roughly a relative improvement of 25\%.

As in a previous work on bulk liquid $^3$He, we have used the 
RN technique to get some insight on the quality of the upper bound 
provided by the FN method. Although the RN method is unstable in the 
sense that the subjacent bosonic component asymptotically overwhelms the 
Fermi signal, the initial slope of the energy with the released time  
$t_{\text{r}}$  can be determined with enough 
precision to make a comparison between different model nodal 
surfaces possible~\cite{casu}.
This kind of analysis is shown in Fig. 2, where the RN energies 
calculated with a  $\psi_{\text{T}}$  containing or not backflow 
correlations are plotted as a function of  $t_{\text{r}}$. The 
Figure shows a reduction of the slope by a factor of two when 
backflow correlations are present in  $\psi_{\text{T}}$  pointing 
to a significant improvement of the upper bound (an exact wave 
function would show zero slope). However, this improvement is 
worse than the one we observed in the past in a similar calculation of 
bulk  $^3$He~\cite{casu}. In fact, also the variance observed in the 
present simulation of the drops is larger than the one 
estimated in bulk. Both comparisons point to a somehow 
incomplete treatment of the inhomogeneity inherent to 
drops. Possible improvements on this line could be 
obtained by changing the functional dependence of 
correlations, from $(|\bf{r}_i-\bf{r}_j|)$ to $(\bf{r}_i,\bf{r}_j)$~\cite{kro4he}. 
Notwithstanding, this 
modification would introduce additional complexity in 
the calculation and require from new correlation 
functions which at present are not very well known.

\begin{figure}
\centering
        \includegraphics[width=0.8\linewidth]{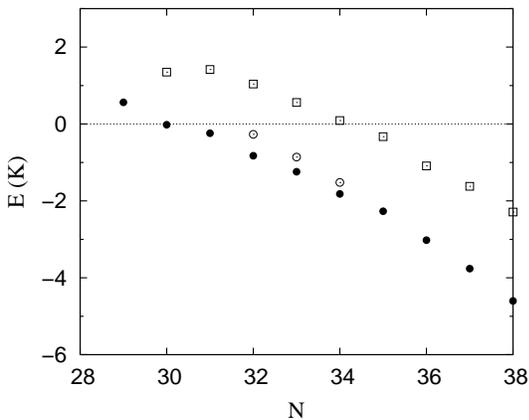}%
        \caption{Threshold limit for self-binding in $^3$He drops. Solid
	circles correspond to the present results. Open circles stand for the
	FN-DMC calculation from Ref. \protect\cite{guardio_dmc}, 
	and squares for the VMC
	results from Ref. \protect\cite{guardio_vmc}.
	}
\end{figure}

The present DMC results for $^3$He drops, near the threshold limit for
self-binding, are compared with recent MC data in Fig. 3. In the Figure,
the VMC energies correspond to the best variational calculation up to
date~\cite{guardio_vmc}. The high quality of the upper bounds there
achieved comes from the introduction of configuration-interaction
correlations in the Jastrow part. The resulting limit for self-binding was
$N=35$, quite close to our result ($N=30$), and the difference with respect
to the present results increases slightly with $N$. The orbitals we have
used, which are essentially the same used in this VMC estimation, do not
have in general good angular momentum quantum numbers. In principle, that
can be considered a defficiency of the model and the results obtained could
be worse than the ones obtained with better wave functions. However,
Guardiola~\cite{guardio_mom} proved using VMC that the energies of $^3$He
drops are independent of the orbital angular moment and that they depend
only on the spin.       

Recently, Guardiola and Navarro~\cite{guardio_dmc} have reported FN-DMC
results for $^3$He drops in the range $N=31-34$ (also shown in Fig. 3).
They use the same interatomic potential and the same Jastrow factor but a
different shell structure in the Slater determinant, and also a different
form for the backflow function. Their results show a minimal number $N=32$
for a bound drop and energies that are above our results in the number
range considered. We have verified that a significant part of the
difference between  that calculation and the present one lies in the different
functional form used for backflow correlations. In Ref. \cite{guardio_dmc},
the form
$\eta(r)= \lambda/r^3$ instead of a Gaussian is used; for a $N=34$ drop it supposes
an energy loss of $\sim 0.26$ K. 
       
In conclusion, we have carried out an accurate calculation of small $^3$He
drops using FN- and RN-DMC that has allowed for a reduction in the 
threshold limit for self-binding; our results show that this minimum number
is $N=30$. Similarly to previous density-functional~\cite{barranco} and
microscopic calculations~\cite{guardio_vmc}, our results confirm that the
ground-state energy is achieved for maximum spin or, in other terms, for
shell configurations where the number of atoms of one of the two species
(up or down) closes a shell. In the $N$ range studied, the energy shows
kinks for $N=30$ and $N=40$ which correspond to magic numbers of the
underlying shell model; its signal is however quite depressed by the
relevance of $^3$He-$^3$He dynamical correlations.

We thank J. Navarro and R. Guardiola for useful
discussions. Partial financial support from DGI (Spain) Grant No. BFM2002-00466
and Generalitat de Catalunya Grant No. 2001SGR-00222 is gratefully
acknowledged. E. S. acknowledges support from MECD (Spain).

\end{document}